# Steels and Stainless Steels

*Stefano Sgobba*
CERN, Geneva, Switzerland

**Abstract**
Steels, and in particular stainless steels, play a crucial role in the construction of large particle accelerators and high-energy physics experiments, of fusion reactors and their superconducting magnet structures. Such projects face severe material challenges, as they require a wide application of tightly specified steel products and grades, featuring a controlled microstructure and adequate mechanical, physical, magnetic and vacuum properties over a wide temperature range. A broad spectrum of relevant examples is presented, issued from the experience maturated within decades of building of large vacuum, cryogenic and associated structural systems that must guarantee a reliable, long-lasting service with limited interventions. The requirements, and in turn the metallurgical processes applied to achieve the final stringent properties are discussed - dictated by mechanical, magnetic or vacuum compatibility constraints and often by a combination of them. Case studies are developed. The study of a few major failure analysis cases and their root causes is also addressed.

## 1   Introduction

Approximately five thousand years before Christ, mankind made a significant advancement by learning to extract metals from their mineral ores—essentially the stones in which these metals were encased. This pivotal development marked the transition from the Stone Age to the Metal Age. The first metal to be successfully smelted was copper, followed by the alloys of bronze and by iron. Copper ores are notably easier to smelt compared to iron ores. The melting point of pure iron is 1538 °C, which is considerably higher than that of copper (1085 °C), lead (328 °C), or tin (232 °C). Although the melting point of iron can be significantly lowered when alloyed with carbon, as it is the case for cast iron, achieving the necessary temperatures to produce liquid iron still requires reaching levels well above 1100 °C. This posed a considerable challenge, as the primitive smelting furnaces used for processing lower melting point metals were not enabling to attain such high temperatures. Consequently, this necessitated a substantial evolution in smelting technology to facilitate the extraction of iron [1].

Iron was the heaven mineral among the ancient Egyptians; the Greeks referred to it as σίδηρος (síderos). This celestial mineral, believed to descend from the heavens ("sīdus" in Latin translates to "star"), was regarded as a divine gift. Indeed, meteorites, which contain an alloy of iron and nickel, were particularly valued because they did not necessitate the complex, labour-intensive, and energy-consuming smelting processes associated with iron extraction and smelting. Consequently, the metallurgy of iron and steel is still referred today to as siderurgy, the science of the heaven, a term that reflects its celestial origins and significance.

Very painful labour conditions in metallurgy plants were maintained over centuries. An engraving contained in an ancient book about Belgium depicts a rolling mill of the Montigny-sur-Sambre plant in 1888 (Fig. 1).

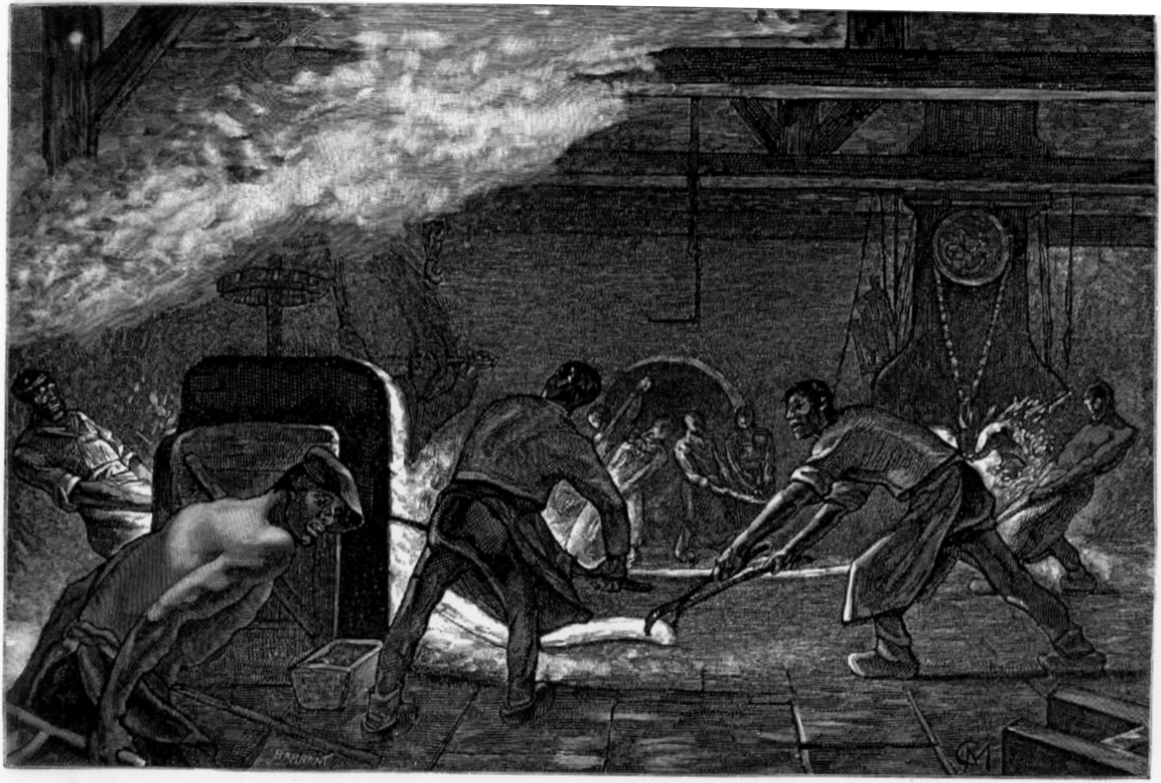

**Fig. 1:** Rolling mill in 1888. Reproduction of the so called « iron and fire » world by Lemonnier (from [2]).

## 2 Steels and cast iron

Depending on the source, steel is defined as either "an iron-based alloy containing carbon, silicon, manganese and other elements" or "an alloy of iron and carbon", with the amount of carbon determining whether the steel is hard or tough. The C content of steel usually falls within a range of 0.3–1.5% by volume. However, any classification of ferrous alloys includes many sub-families and subcategories, such as low carbon, ultra-low carbon (with only a few ppm of carbon), medium-high carbon, tool steels, plain steels and high alloy steels (with more than 10% alloying elements other than carbon). The number of industrially available steel designations worldwide today is impressive: around 75000 standards and proprietary designations are listed in traditional handbooks such as "Stahlschlüssel" [5]. Some online databases are also very comprehensive, encompassing both actual and historical standard and proprietary designations. A notable example is Total Materia [6].

As steels are defined as an alloy of iron and carbon, their understanding is based on the iron-carbon diagram, which is generally limited by pure iron on the left and by cementite ($Fe_3C$) on the right. In spite of the metastability of this phase, it is representative of an upper limit in the diagram. As demonstrated in the diagram in Fig. 2, the steels (up to approximately 2% carbon) and the cast irons, which are higher in carbon (between 3 % and 4.5%), can be readily identified. The composition of cast iron is centred around a eutectic point, enabling melting and casting temperatures to be much lower than those of steels. In the nascent stages of iron metallurgy, the primary focus of smelting processes was on C-rich compositions, with subsequent processing through forging or milling. The Damascus blade is constituted of hypereutectoid steel (1.44% C) with a high phosphorus content (0.19% P) [7]. This approach played a crucial role in the development of iron-based technologies.

Notwithstanding the pivotal role of innovative materials in contemporary technologies, steel consumption has been progressively rising for several decades. In 1958, global production was recorded at 270 million tonnes. By 1975, this figure had increased to 644 million, and by 2024 it had further risen to 1.885 billion [8]. Cast iron blocks are also utilised at CERN in the context of the Radiation to Electronics (R2E) project, which is aimed at mitigating radiation effects at both the circuit and system level by means of shielding the LHC's electronics against radiation. These blocks were produced in Belgium by a company that owns a state-of-the-art automatic casting line (Fig. 3). Evidence has been presented which indicates that conventional materials retain their significance for high-tech applications. Moreover, their production can benefit from contemporary processing methodologies.

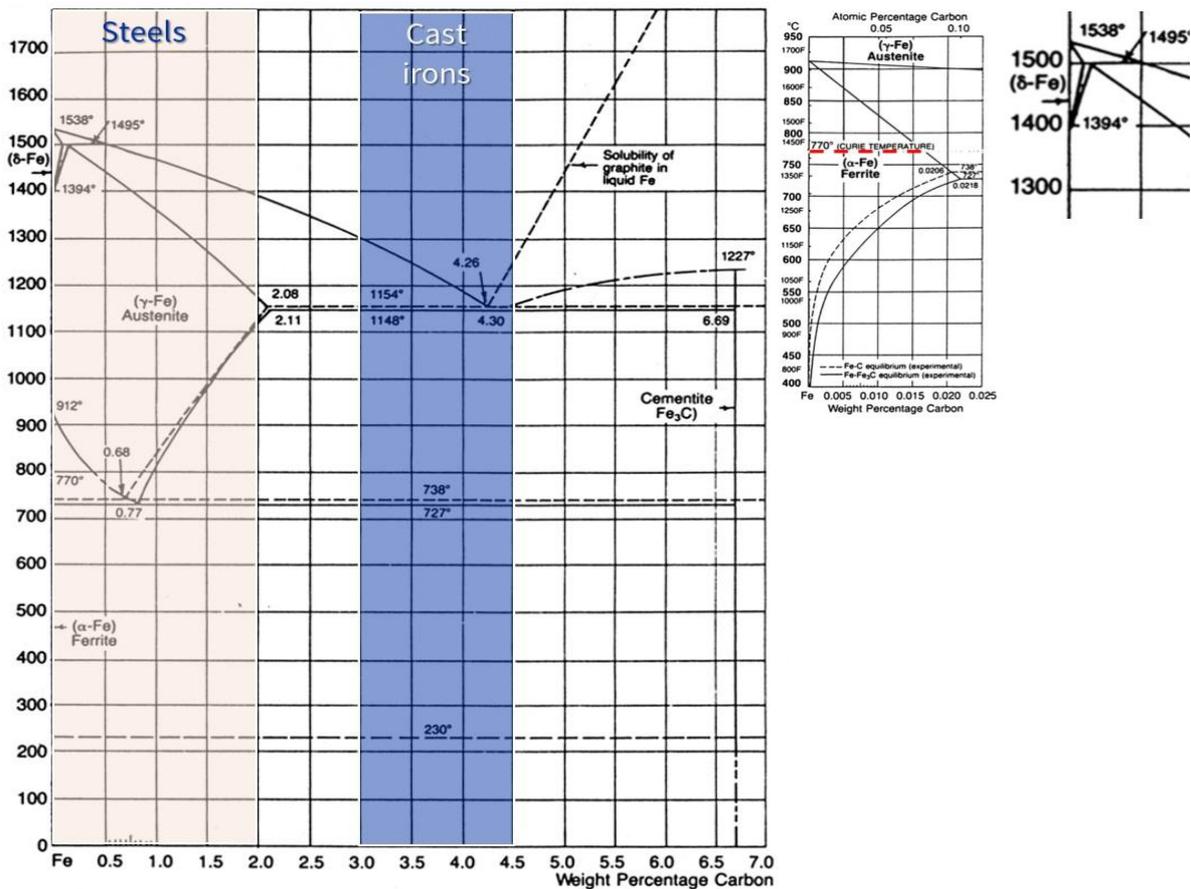

**Fig. 2:** Portion of the Fe-C diagram (limited on the right by cementite) depicting the range of existence of steels and cast irons. Details of the α- and δ-ferrite domains are also illustrated [9].

When predicting the phases starting from the Fe-C diagram, it looks evident that ferrite exists in two temperature ranges, low and high temperature, respectively called α- and δ-ferrite. The existence range of α-ferrite is limited to a very few amounts of carbon, some 200 ppm. The ferrite lattice is body centred cubic (bcc): it is in principle a quite loose lattice, however the interstitial sites are small, hence it is not able to accommodate a significant amount of carbon. Ferrite is ferromagnetic under the Curie temperature (770 °C). Ferritic steels feature a Ductile to Brittle Transition Temperature (DBTT), hence due to the limited ductility and toughness under the DBTT they are not used for cryogenic applications, with a few exceptions such as the yoke laminations of superconducting magnets [10]. The yield strength of ferritic steels is very well defined, as they possess a distinct upper and lower yield strength with transient oscillation of stress between the two. This behaviour is detrimental for the formability, causing uncontrolled deformation and undesired micro-necking and for the surface finish after forming. Typical microstructures featuring ferrite are shown in Fig. 4.

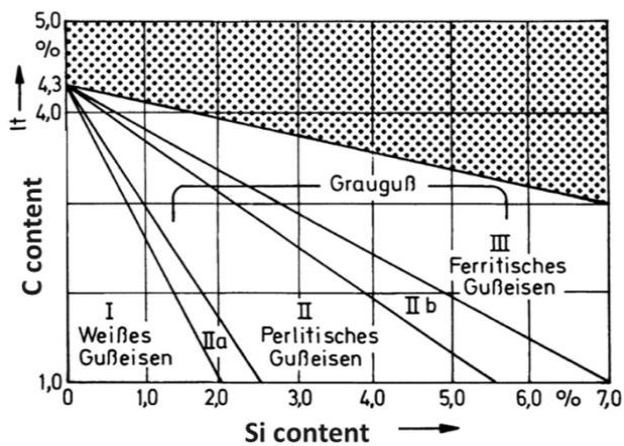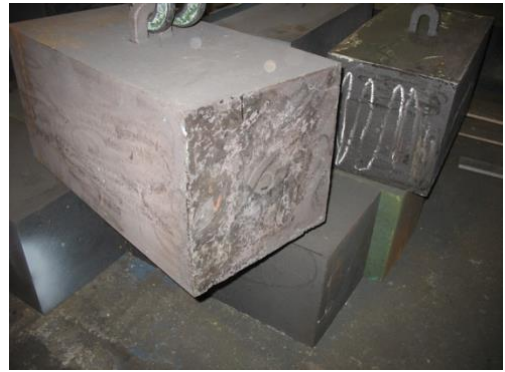

**Fig. 3:** Left - The Maurer diagram, allowing the microstructure of cast irons to be predicted from their C and Si content (source: [11]). Developed by E. Maurer in 1924, the diagram shows the influence of C and Si content on the microstructure of cast irons (assessed from 30 mm bars). The three main fields represent white (I), pearlitic (II) and ferritic cast iron (III), respectively. Transitional zones IIa and IIb are for malleable and pearlitic-ferritic cast iron, respectively. Higher C and lower Si favour cementite and white cast iron structures. Higher Si content promotes graphite formation during slow cooling, leading to a grey cast iron microstructure. Right - Spheroid-graphite cast iron shielding blocks for the LHC-R2E project. 436 t procured from NV Ferromatrix/BE.

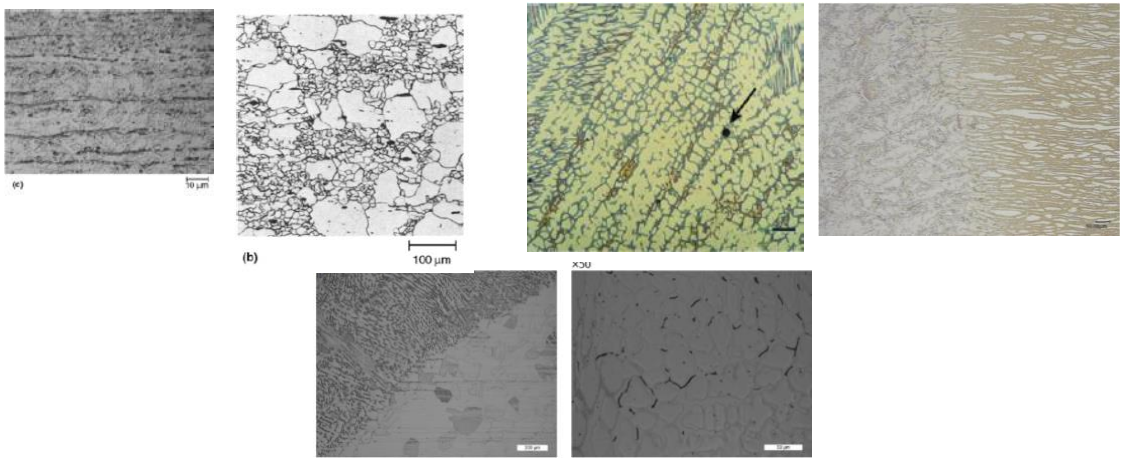

**Fig. 4:** Examples of microstructures containing ferrite or fully ferritic microstructures. From left to right and top to bottom: δ-ferrite stringers (dark phase) in 17-7 PH stainless steel after hot and cold processing (from [12]); full α-ferrite structure in a 0.013% C ferritic steel, finish rolled at 845 °C and coiled at 725 °C (ibidem); δ-ferrite (dark phase) at the dendritic boundaries in austenitic stainless steel welds (ibidem); δ-ferrite (light grey phase) in a URANUS 45® duplex stainless steel and its welds (beam dump at LHC Point 6); δ-ferrite (dark phase) under the form of stringers and as precipitated in a LHC repair weld of a LHC austenitic stainless steel diode box.

Conversely, the austenitic phase, which spans a considerably broader range within the phase diagram, is of paramount importance for cryogenic applications. Austenite (also known as γ-phase) is face-centred cubic (fcc). It is very compact but features very large interstitial sites which have the capacity to accommodate a higher quantity of carbon. It is paramagnetic, does not undergo a ductile to brittle transition and has high ductility and toughness. The phase does not exhibit a distinct yield point; consequently, a yield strength of 0.2% is defined in a conventional manner. Austenite is high formable due to its ductility. The absence of discontinuous yielding is an advantage.

Other phases are specific to ordinary steels, such as pearlite. During the transformation at the eutectic point, starting from austenite of the eutectic composition, the microstructure forms through the separation of ferrite and cementite. This results in the characteristic alternating lamellar pattern. Although pearlite is generally not visible at the macroscopic scale, it can be observed in exceptional examples of craftsmanship such as the Damascus swords mentioned above, which have a coarse pearlite microstructure together with proeutectoid cementite (Fig. 5). Indeed, moving away from the eutectic composition, excess ferrite or cementite can form, known as proeutectoid phases. Fully pearlitic steels are of limited practical interest because, while they exhibit high strength and hardness, they tend to have low ductility. Typically, pearlite appears in combination with other microstructures rather than in pure form.

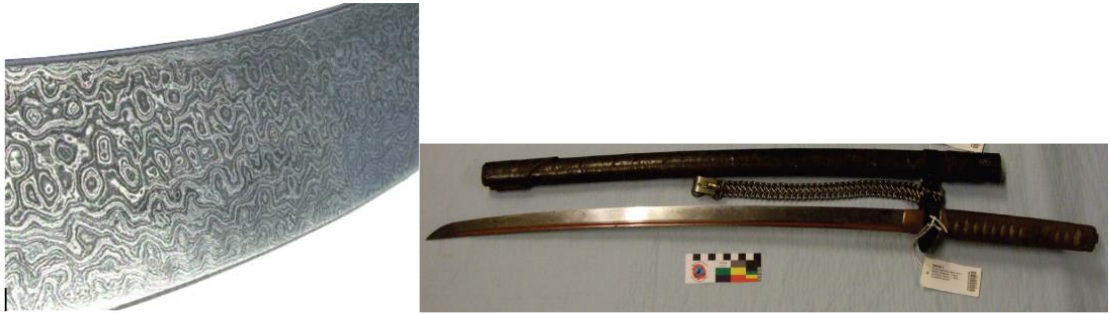

**Fig. 5**: Left - "True" Damascene blade macrostructure. The real Damascene pattern (also called "water pattern") must come from a striated precipitation of $Fe_3C$ particles and not from folding and welding. Right - A Katana sword (from [13]).

Bainite is a microstructure that forms when steel cools from the austenitic phase at intermediate cooling rates. It consists of a fine, alternating mixture of ferrite and cementite, often so minute that it can only be seen with transmission electron microscopy. This microstructure imparts a unique combination of strength, toughness, and ductility. Historically, bainitic steels are exemplified by traditional Japanese samurai swords, known as "Katana," which showcase this intricate microstructure. The bainitic microstructure contributes to the sword's remarkable balance of sharpness, bendability and shock absorption. Some Katana steels are well designated (such as L6 grade). Modern bainitic steels are also used in various industrial applications due to these properties.

For even faster cooling, austenite transforms directly into martensite. This transformation is of paramount importance for steels used in practical applications: the so-called "heat-treatable steels" are primarily steels that develop a martensitic phase. These steels are easy to quench, unlike constructional steels, which require very high cooling rates—sometimes as rapid as one second—to form martensite. Such rapid cooling is often incompatible with thick products that cannot be quenched through their entire core, leading to the formation of bainite or pearlite in place. Consequently, constructional steels are considered non-heat-treatable (Fig. 6 top). In contrast, steels that can be air-quenched form martensite even at much lower cooling rates (Fig. 6 bottom). This means that heavy gauge products in these steels can develop martensite throughout their volume, even if the bulk cooling rate is slower than at the surface.

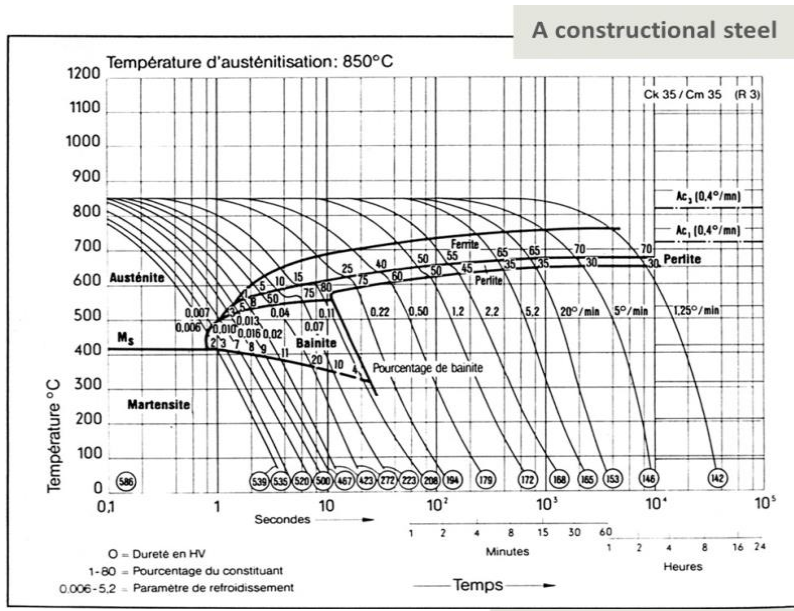
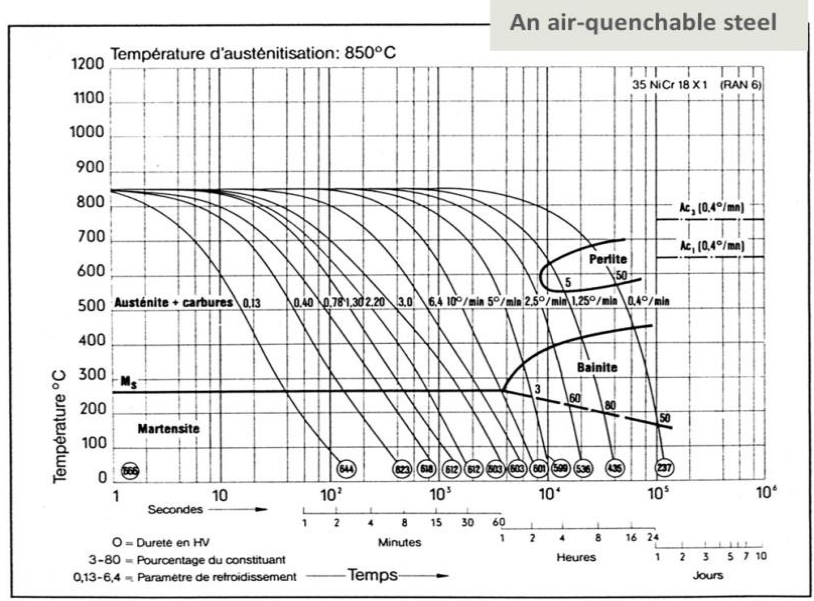

**Fig. 6**: Top - Continuous Cooling Transformation (CCT) phase diagram of a constructional steel. The martensitic range is very limited and can only be reached with a very high cooling rate, which during quenching is only achievable on the surface of the product. Bottom - CCT of a heat-treatable steel. The martensitic range is much more extensive, and can be reached even with moderate cooling rates (e.g. 10 °C/min).

Martensite results from the deformation of the body-centred cubic (bcc) lattice into a tetragonal structure. This distortion, which imparts hardness, is proportional to the steel's carbon content. The ability to form martensite depends largely on the steel's chemical composition.

Carbon atoms occupy specific positions within the tetrahedral sites of the microstructure. The resulting microstructure contains martensite laths, which are characteristic signatures of martensitic transformation. Martensite can also be identified through its physical properties: it is a ferromagnetic phase, detectable via a magnetic permeameter, or at the microscale using Magnetic Force Microscopy (MFM). Additionally, Electron Backscatter Diffraction (EBSD) allows for the identification of the martensitic lattice in contrast to the austenitic phase. An as-quenched, heat-treatable steel with a fully martensitic structure exhibits exceptional hardness, up to 800 HV to 900 HV. However, this high hardness makes the steel extremely brittle with very low fracture toughness. To mitigate these drawbacks, an additional heat treatment called tempering is usually performed. Tempering partially reverts some of the martensite to softer phases, striking a balance between strength and toughness. A tempered martensite consists of a mixture of reverted ferrite and cementite. The hardness decreases accordingly, resulting in improved ductility, reduced area reduction, and increased toughness.

In German, heat-treatable steels are referred to as *Vergütungsstähle*; in French, as *Aciers d'amélioration*; and in Italian, as *Acciai da bonifica*. All these terms emphasise that their properties can be significantly improved through the tempering process.

The quenchability of steels can be evaluated using the so-called Jominy test. Developed by Walter Jominy, employed by the Westinghouse Electric & Manufacturing Company in the United States at the beginning of the 20$^{th}$ century, this test determines whether a steel is "improvable" through heat treatment. In the test, a steel bar heated to a specific temperature and extracted from the furnace is mounted on a fixture. A water jet is directed at one end of the bar, cooling it from that end. The steel's quenchability is then assessed by measuring the hardness profile along the length of the bar, from the quenched end outward. A fully quenchable steel will exhibit a high and nearly uniform hardness across a significant portion of the bar, resulting in a flat hardness profile. In contrast, a constructional steel will show a sharp decrease in hardness away from the quenched end, indicating limited hardenability.

It is important to note that martensite can only form from austenite. Without a reversion to austenite, heat treatments to alter the microstructure are not possible. During cooling from austenite, different microstructures develop depending on the cooling rate: slow cooling produces pearlite (coarse or fine, depending on the rate), moderate cooling results in bainite, and rapid quenching leads to martensite. Subsequent tempering can then convert martensite into tempered martensite. The hardness of these microstructures varies: coarse pearlite is softer, fine pearlite is harder, bainite is harder still, tempered martensite is even harder, and as-quenched martensite is the hardest. However, as mentioned earlier, untempered martensite is extremely brittle and not suitable for structural use.

## 3    Stainless steels

Stainless steel plays a crucial role in the construction of modern accelerators. The example of the dipole and quadrupole magnets of LHC is representative in this respect. A special austenitic stainless steel has been developed [14] for the beam screen and the cooling capillaries of the machine vacuum system, retaining high strength, ductility and low magnetic susceptibility at the working temperature between 10 K and 20 K. Several tens of kilometres of components have been produced in this special grade. The magnet cold bore was manufactured as a seamless 316LN tube. 316LN is also the reference grade for the shrinking cylinder of the dipole magnets (2500 bent plates of 15.35 m of length, 10 mm thick for a total weight of approx. 3000 t) The plates were welded longitudinally by a special Surface Tension Transfer (STT) technique combined with traditional pulsed MIG welding. More than 2800 magnet end covers of complex shape, including several nozzles, have been fabricated starting from HIPed 316LN powders and near net shaped into geometry close to the final form [15]. Between the magnets, some 1600 interconnections consist of several thousand of leak tight components to be integrated, mainly working at cryogenic temperature (1.9 K). Interconnection components are also mainly based on austenitic stainless steels. For the convolutions of the several thousands of bellows involved in the machine and working under cyclic load at 1.9 K, a special remelted 316L grade has been selected showing an extremely low inclusion content and improved austenite stability at the working temperature. This grade is highly formable [16].

## 3.1 Families of stainless steels

Stainless steel can be defined as a ferrous alloy containing a minimum of 12 % Cr [17], with or without other elements [18,19]. Chromium, as well as additional alloying elements, imparts corrosion and oxidation resistance to steel. Figure 7 is an iron-chromium diagram, which is the foundation of stainless steels [17]. On the 100% Fe axis of the diagram, one recognises the stability domains of the various phases of iron as a function of temperature: α- and δ-iron, corresponding to the ferromagnetic ferritic phase of body centred cubic (bcc) structure, which are present up to 912 °C and in the ranges between 1394 °C and 1538 °C, respectively, and γ-iron, corresponding to the austenitic phase of face centred cubic structure (fcc), in the range between 912 °C and 1394 °C. This phase is non-ferromagnetic. β-iron is an obsolete designation of paramagnetic α-iron above the Curie temperature which is not really a distinct phase.

In the diagram of Fig. 7 [20], the ferritic phase is extensive while the γ phase is limited to a loop. This diagram is the basis for identifying the two first families of stainless steels: ferritic types, with Cr contents between 14.5% and 27%, and martensitic types, that are iron-chromium steels with small additions of C and other alloying elements, usually containing no more than 14% Cr (excepting some types with 16% to 18% Cr) and sufficient C to promote hardening. From the diagram it can be seen that ferritic grades do not transform to any other phase up to the melting point, and hence can only be strengthened by cold working. Martensitic grades, due to the addition of C to the Fe-Cr system enlarging the domain of stability of the γ-phase, can be hardened by a rapid cooling in air or a liquid medium from above a critical temperature. This results in grades that have excellent strength. Finally, duplex stainless steels having a dual-phase microstructure of austenite and ferrite are applied where mechanical properties have to be guaranteed in an intermediate temperature range and ferromagnetism is not a concern. They are one of the candidate materials at CERN for structural components of beam dumps.

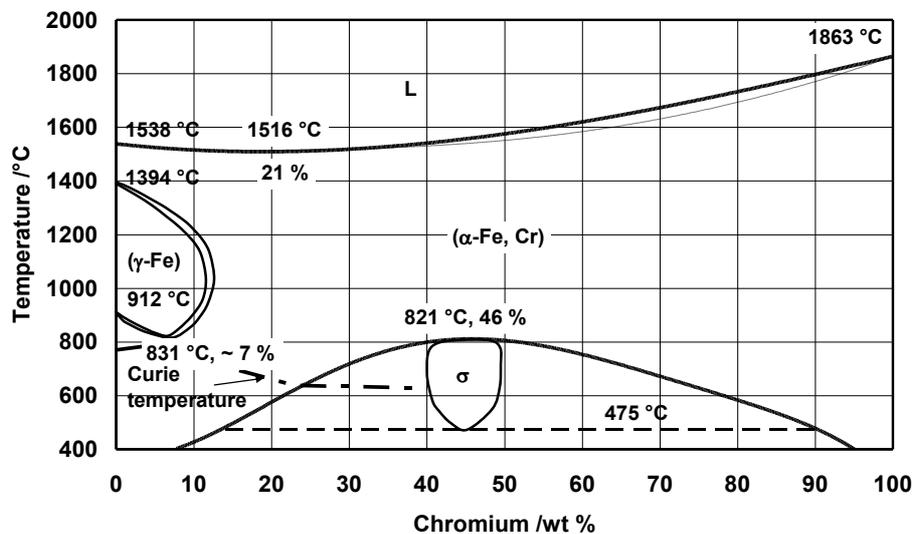

**Fig. 7:** The Fe-Cr phase diagram (from [20])

Ferritic and martensitic grades are generally inferior, in terms of corrosion resistance, to austenitic grades. They are ferromagnetic and undergo ductile-to brittle transition at cryogenic temperatures. Moreover, at temperatures in the range of 1000 °C or more (solution annealing), ferritic grades undergo grain growth, while martensitic grades are subject to a loss of the martensitic hardening phase due to a reversion of martensite to austenite in the annealed condition associated with a high temperature treatment. On the other hand, austenite is non-magnetic, does not undergo any ductile-to-brittle transition below RT, and is less subject to grain growth during vacuum firing. For these reasons, the austenitic grades are a first choice for vacuum applications. Austenitic grades, due to their high Cr and Ni content, are the most resistant to corrosion of the stainless steel family [17]. Because of their limited application in vacuum devices, precipitation-hardening grades are not mentioned here.

## 3.2 Austenitic stainless steels

These are formed by the addition of elements (Ni, Mn, N…), to the Fe-Cr system of Fig. 7, broadening the domain and enhancing the stability of the γ phase. For sufficient additions of a balanced amount of adequate alloying elements, the formation of ferrite can be suppressed and the tendency to form martensite on cooling or during work hardening can be partially or completely suppressed. As an example, the addition of 8% to 10% Ni to a low C FeCr steel can already allow a relatively stable austenitic structure at RT to be obtained. The '304-type' stainless steel (18Cr8Ni, on an Fe basis) is a typical example of a very common iron-chromium-nickel austenitic stainless steel. This grade is generally applied in its low (L) carbon version 304L (C ≤ 0.030%), showing enhanced corrosion resistance and ductility especially in welded structures. Figure 8 is a pseudobinary section of the FeCrNi ternary diagram for increasing Cr+Ni contents [21]. In this diagram, for a total Fe of 70%, it is straightforward to visualise the basic 304 type of austenitic stainless steel. Continuous lines in the diagram correspond to real transformations, dashed lines to transformations not observed in practical conditions. For a 20% Cr content and 10% of Ni (composition corresponding to the far left dashed vertical line), the diagram shows that cooling from the usual solution annealing temperature of 1050 °C, can result in some residual ferrite. Indeed, 304L grade retains, when cooled from solution annealing (or precipitates during a welding operation), a given amount of untransformed δ-ferrite. Increasing Ni and reducing Cr (vertical lines to the right) enhances stability of γ-phase against precipitation of ferrite and, for sufficient Ni, can totally suppress it.

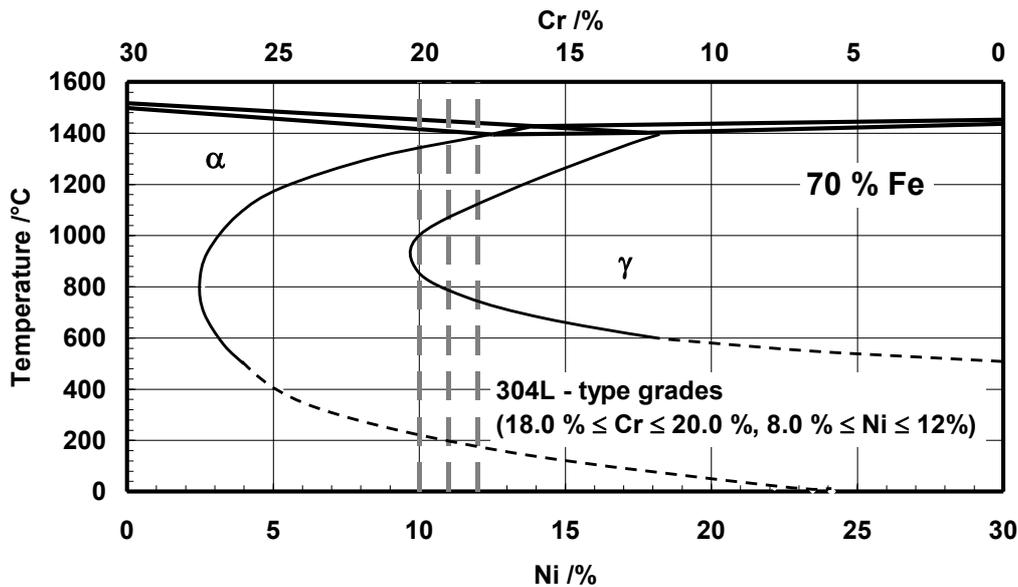

**Fig. 8:** Cross-section of the Fe-Cr-Ni ternary phase diagram [21]

From the diagram in Fig. 8, it appears clear that:

1) austenitic stainless steels may contain residual amounts of δ-ferrite, which can be critical due to its reduced toughness and to its ferromagnetic nature for specific vacuum applications, particularly for components that have to be applied at cryogenic temperature and/or in critical magnetic environments;

2) elements like Ni extend the domain of austenite, while Cr reduces it. Elements such as C, N, Mn (up to some extent)… or a combination of them play a role similar to Ni ('Ni-equivalents', $Ni_{eq}$), while elements like Mo, Si, Nb… or their combination act as 'Cr-equivalents' ($Cr_{eq}$). Schaeffler, DeLong, Hull, or Espy diagrams allow the total ferrite formation effects to be predicted in a multielement system, such an industrial stainless steel grade and its welds (Section 3.5.1).

## 3.3 Grades of practical interest for vacuum applications

The first four rows in Table 1 show the composition of the grades of stainless steels (304L, 316L, 316LN) of main interest for vacuum applications and covered by CERN specifications. The mentioned composition ranges are based on the last version of the CERN specifications [22−29], with reference to the relevant standard grade designations.

**Table 1:** Grades of practical interest for vacuum, cryogenic and/or structural applications. Composition ranges according to CERN specifications, wherever applicable, or EN 10088 [26]. Single values are maximum admitted values.

| Grade (AISI, or 'commercial designation') | Grade (EN, symbolic and numeric) | C | Cr | Ni | Mo | Si | Mn | N | Others |
|---|---|---|---|---|---|---|---|---|---|
| 304L [22,23] | X2CrNi19-11 1.4306 | 0.030 | 17.00−20.00 | 10.00−12.50 | | 1.00 | 2.00 | | P≤0.030, S≤0.015, Co≤0.10 |
| 316L [24,25] | X2CrNiMo18-14-3 1.4435 | 0.030 | 17.00−19.00 | 12.50−15.00 | 2.50−3.00 | 1.00 | 2.00 | | P≤0.030, S≤0.015, Co≤0.10 |
| 316L for bellows [26] | X2CrNiMo18-14-3 1.4435 | 0.030 | 16.50−19.00 | 13.50−16.00 | 2.50−3.00 | 1.00 | 2.00 | 0.050 | P≤0.030, S≤0.010, Co≤0.10 |
| 316LN [27−29] | X2CrNiMoN17-13-3 1.4429 | 0.030 | 16.00−18.50 | 12.00−14.00 | 2.00−3.00 | 1.00 | 2.00 | 0.14−0.20 | P≤0.045, S≤0.015, Co≤0.10 |
| 316Ti [30] | X6CrNiMoTi17-12-2 1.4571 | 0.08 | 16.50−18.50 | 10.50−13.50 | 2.00−2.50 | 1.00 | 2.00 | | P≤0.045, S≤0.015, 5xC≤Ti≤0.70 |
| 'P506' [14] | - | 0.030 | 19.00−19.50 | 10.70−11.30 | 0.80−1.00 | 0.50 | 11.80−12.40 | 0.30−0.35 | P≤0.020, S≤0.002, B≤0.002, Cu≤0.15, Co≤0.10 |
| FXM-19 ('Nitronic®50') [31] | X2CrNiMnMoNNb21-16-5-3 1.3964 | 0.060 (0.030[1]) | 20.50-23.50 | 11.50-13.50 | 1.50-3.00 | 1.00 | 4.00-6.00 | 0.20-0.40 | 0.10≤Nb≤0.30, 0.10≤V≤0.30, P≤0.045, S≤0.030 |

AISI 304L is a general purpose grade. For vacuum applications, the grade should be purchased through a careful specification, aimed at achieving a substantially austenitic microstructure and a controlled maximum level of inclusions (Section 3.4). Even in its higher alloy version 1.4306 specified by CERN, due to the limited amount of alloying elements, its magnetic susceptibility can be subject to increase by martensitic transformation. This can occur upon cooling to RT or to cryogenic temperatures, and following work hardening at RT or lower [32]. Its selection should be reconsidered in applications where an increase in magnetic susceptibility might be of concern. The price of this grade is approximately 5 EUR/kg (at mid-2025 rates) for a general purpose version, and 10 EUR/kg for a vacuum specified wrought product.

---

[1] ITER requirement

AISI 316L is a Mo bearing grade. Mo enhances corrosion resistance and austenitic stability versus martensitic transformation. However, the ferrite-promoting characteristics of Mo (see Sections 3.2 and 3.5.1.1) mean that adjustments to the levels of Cr and Ni are required to achieve an almost fully austenitic microstructure. Due to its formability, ductility, and increased austenitic stability compared to 304L, this grade is covered by a special CERN specification applicable to the material of the bellows convolutions of the LHC interconnections. In the form of thin sheet for these convolutions, prices rise up to 50−80 EUR/kg. The high alloy EN 1.4435 version is preferred to EN 1.4404 for vacuum applications, featuring a composition more stable against martensitic transformations and in general also against δ-ferrite precipitation in welds. Wrought 316L products for vacuum applications, newly covered by CERN specifications [24,25], have prices in between 304L and 316LN for equivalent steelmaking and metalworking processes and inspections.

AISI 316LN is a nitrogen bearing stainless steel. N increases austenite stability against martensitic transformations and is a powerful austenite former with respect to ferrite. N substantially increases strength, while allowing ductility to be maintained down to cryogenic temperatures [28]. Due to limited softening compared to 304L and 316L, 316LN is the grade preferentially selected when vacuum firing is required. Prices in the range of 20 EUR/kg (bars) to 35 EUR/kg (plates) or are common for wrought products issued from Electroslag Remelted (ESR) ingots. At the level of ingot remelting and for steelmaking companies equipped with last generation, highly productive ESR units, the additional cost of ESR applied to austenitic stainless steel grades can be very limited, of the order of 1 EUR/kg at the ingot level (2015 rates [33]). However, additional value is brought by redundant multidirectional forging of the ingots. Open die forged ESR 316LN products (Section 3.4) may have prices well above 100 EUR/kg.

AISI 316Ti is an example of a Ti 'stabilised' grade. Stabilised grades contain higher C than low C grades (limited to 0.030% C max). They are alloyed with Nb, Ta or Ti to prevent carbide precipitation. AISI 316Ti, or similar stabilised grades (AISI 321, AISI 347), are offered by steel suppliers or component producers as an alternative to low carbon grades. Stabilised grades should be generally avoided for demanding vacuum applications, since the addition of the stabiliser elements results in precipitate carbides, reducing the cleanliness of the steel (Section 3.4) and toughness in specific low temperature applications.

P506 is a stainless steel specially developed by CERN [14], belonging to the family of high Mn, high N stainless steels and successfully produced by Böhler (now Voestalpine Böhler Edelstahl GmbH)/AT and Aubert et Duval/FR (under a different designation). This special composition allows low relative magnetic permeability (<1.005) to be maintained down to cryogenic temperatures in the base material and in its welds. A full stability of the austenite versus δ-ferrite precipitation in welds, and/or versus martensitic transformations, even when deformed at very low temperatures, is guaranteed by the alloy composition [34]. Its price is comparable to that of multidirectionally forged products in 316LN grade.

FXM-19 is a high strength nitrogen alloyed austenitic stainless steel, also known as S20910 or NITRONIC®50 and widely used in structural applications for fusion magnet systems, including at cryogenic temperature under the form of 3D redundantly forged products [35-38]

## 3.4 Special requirements on the microstructure of stainless steels for vacuum applications

Figure 9 shows an end fitting welded to a thin bellows convolution, which was intended for application in the vacuum system of the Compact Muon Solenoid (CMS) experiment. The bellow ends are machined from stainless steel forged bars, showing strings of non-metallic inclusions aligned parallel to the axis of the bar (the primary metal flow direction). These inclusions result in a leak through the material thickness of some $10^{-4}$ mbar·l·s$^{-1}$ [39].

Several leaks through the base material thickness of flanges of the plug in modules of LHC magnet interconnections have been reported and investigated [40]. In this case, leaks are due to exogenous macroinclusions (Fig. 10). Macroinclusions are understood in terms of the presence of entrapped slag or refractories added to protect the steel melt during the melting phases, segregated during the solidification and not fully removed due to an insufficient top discard of the steel ingot. Slag trapped in the ingot results in elongated macroinclusions during the hot processing of the steel (forging of ingots, rolling of bars and plates, extrusion of shapes).

A complete specification of a stainless steel for vacuum applications must consider the risk of leaks due to the presence of non-metallic micro- and macro-inclusions embedded in the microstructure of the final product. Segregations occurring during primary solidification are also pernicious. Indeed, the direct transformations of the primary melting ingots without a remelting process, such as ESR or Vacuum Arc Remelting (VAR), might not allow the ingot structure to be completely broken and homogenised and can result in unrecrystallised volumes and segregations in the final products [42].

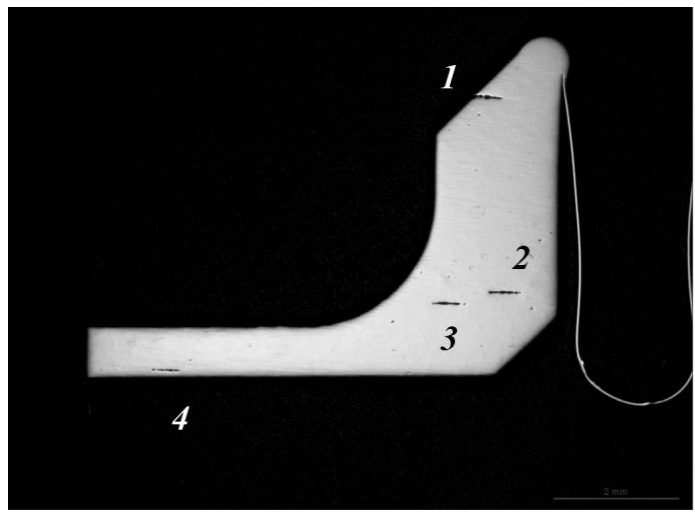

**Fig. 9:** Bellow end fitting machined from an AISI 316LN round bar, showing alignment of oversized (1,2,3) and thick (4) B type inclusions up to class 2, classified according to standard ASTM E45 [41]. CERN specification [29] imposes a class of inclusions at most 1 for type B inclusions and a half-class above this limit in up to 2% of the fields counted.

The risk of leaks can be reduced by one or more of the following actions.
1) In the technical specification, impose a maximum allowed content of non-metallic inclusions. CERN specifications impose maximum limits according to the standards in force.
2) Since microscopical test methods for determining the inclusion content of steel are not intended for assessing exogenous macroinclusions, impose a steelmaking process. Stainless steels are primary melted in an electric furnace and decarburised through an Argon-Oxygen Decarburisation (AOD) or Vacuum-Oxygen Decarburisation (VOD). These processes, consisting of a single melting step, might not guarantee alone a sufficient homogeneity of the ingot. Imposing a remelting process such as VAR or ESR ensures the homogeneity of the material will be effectively influenced. In addition, remelting reduces the impurity and the microinclusion content of the final products, allows high density and a lack of macrosegregations with shrinkage cavities to be achieved [43].

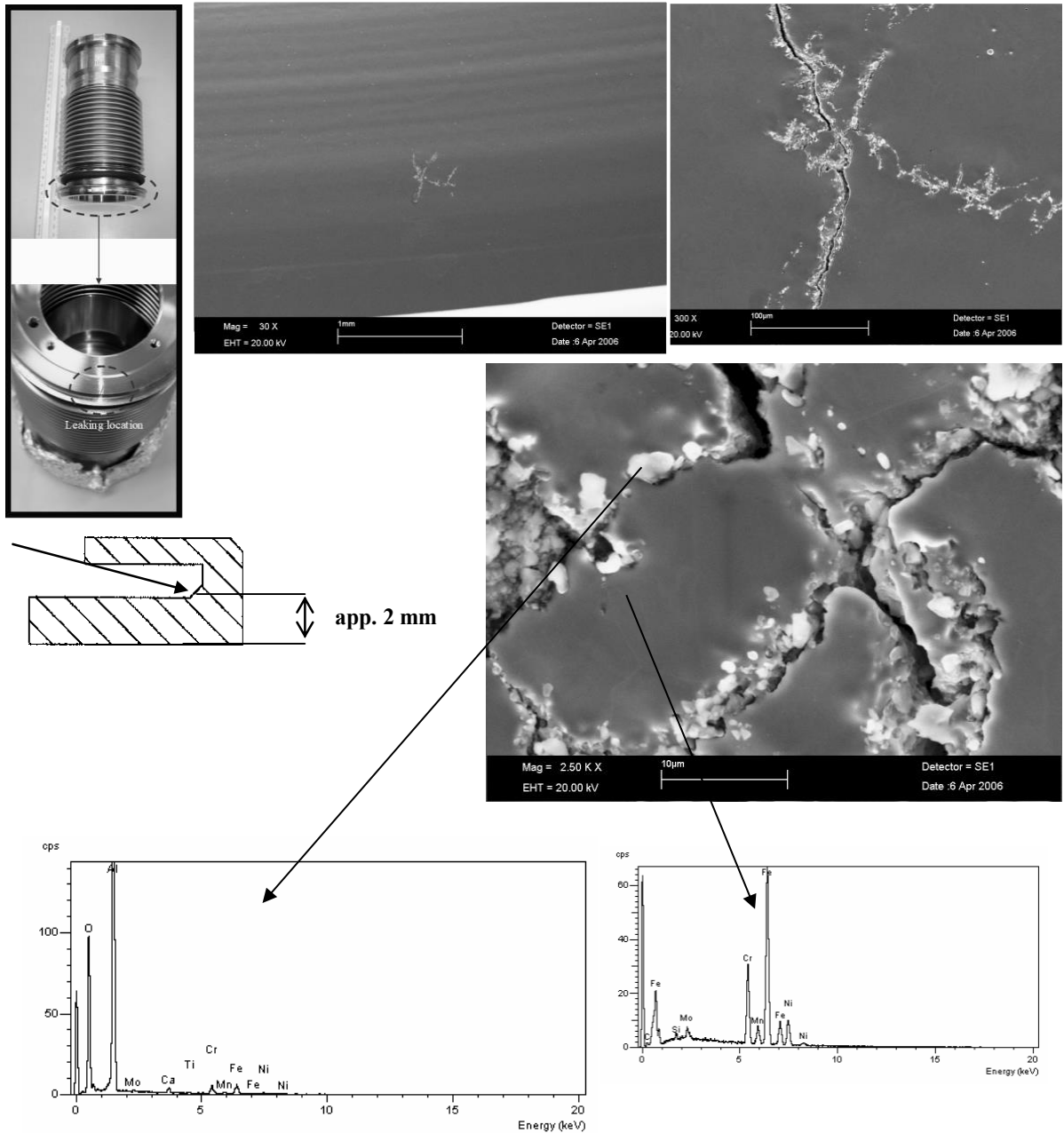

**Fig. 10:** Several plug-in modules of the LHC interconnections have been found leaking at the flange location. Flaws associated with the presence of Al, O, and Ca were detected on the two opposite surfaces of the throats of the cooling exit tube. Al and Ca are typical elements contained in slag or refractories. Entrapped residues of slag resulted in macroinclusions. Elsewhere, cross-sectional micro-optical observations of a non-leaking area only showed B type inclusion up to class 1 (worst field) [32].

3) Specify three-dimensionally, redundantly forged products [28]. Upsetting steps allow the alignment of the inclusions which are elongated by two-dimensional steps of previous forging or rolling to be broken, thus reducing the risk of leak through the thickness.

4) Specify a minimum forging or rolling reduction ratio.

5) Specify products with a fine and homogenous grain size.

6) Impose a final separate solution annealing rather than allowing mill-annealing from a high temperature process, in order to facilitate the achievement of a controlled, homogeneous and non-segregated microstructure [42], more easily inspectable by Non-Destructive Testing (NDT), see point 8 hereafter.

7) Avoid breaking the fibre of the product by machining walls perpendicular to the direction of the material flow, as it was the case in the example of Fig. 9.

8) Introduce NDT such as Ultrasonic Testing (UT) not only on the final product, but also on semifinished products resulting from intermediate steps of steel processing. NDT procedures and acceptance criteria should be the object of an agreement between supplier and customer.

In conclusion, when selecting stainless steels for vacuum applications in accelerators, a mere material designation or the availability of general-purpose stock should never be the deciding factor. Critical parameters must be explicitly specified and strictly controlled. In some cases, this control must extend beyond the final product to the definition and application of a quality control plan involving tests at different stages of the production process.

## 3.5 Aspects related to joining processes of stainless steels in the framework of vacuum applications

### 3.5.1 Precautions during welding

Austenitic stainless steels are readily welded by conventional arc techniques such as Tungsten Inert Gas (TIG), MIG, or beam techniques such as laser and EB. Properly specified and qualified welds according to standards in force[2] can be fully sound. Tensile strengths equal to or higher than the minimum specified for the base metal can generally be obtained, as well as satisfactory fatigue crack growth rates, however for heavy gauge welds and/or specific combinations of parent and filler metals it might be challenging to obtain satisfactory joint strength [49]. Moreover, some additional aspects are relevant for welding austenitic stainless steels in the framework of a vacuum application.

#### 3.5.1.1 Possible presence of δ-ferrite in austenitic stainless steel welds

As mentioned in Section 3.2, austenitic stainless steels can contain residual amounts of δ-ferrite, which can be critical in applications to accelerators due to its reduced toughness and ferromagnetic nature. Constitution diagrams for stainless steel weld metals such as the one presented in Fig. 11 allow the ferrite content to be estimated in the as-deposited weld on the basis of the composition of the base and, when applicable, of the weld filler metal. Control of ferrite to minimum or zero level might be required in the context of specific accelerator applications (cryogenics, components close to the beam). Since fully austenitic stainless steel welds are more sensitive to microfissuring, specific precautions have to be taken such as reducing the weld heat input, minimizing restraint, designing for low constraint, and keeping impurities in the stainless steel composition to minimum levels [19].

---

[2] Welds qualified according to standards ISO 15614-1 [44] for arc welding, ISO 15614-11 [45] for electron and laser beam welding, with reference to standard ISO 6520-1 [46], and level B of ISO 5817 [47] or ISO 13919-1 [48] for classification and quality levels for imperfections, respectively, are generally specified at CERN for vacuum applications.

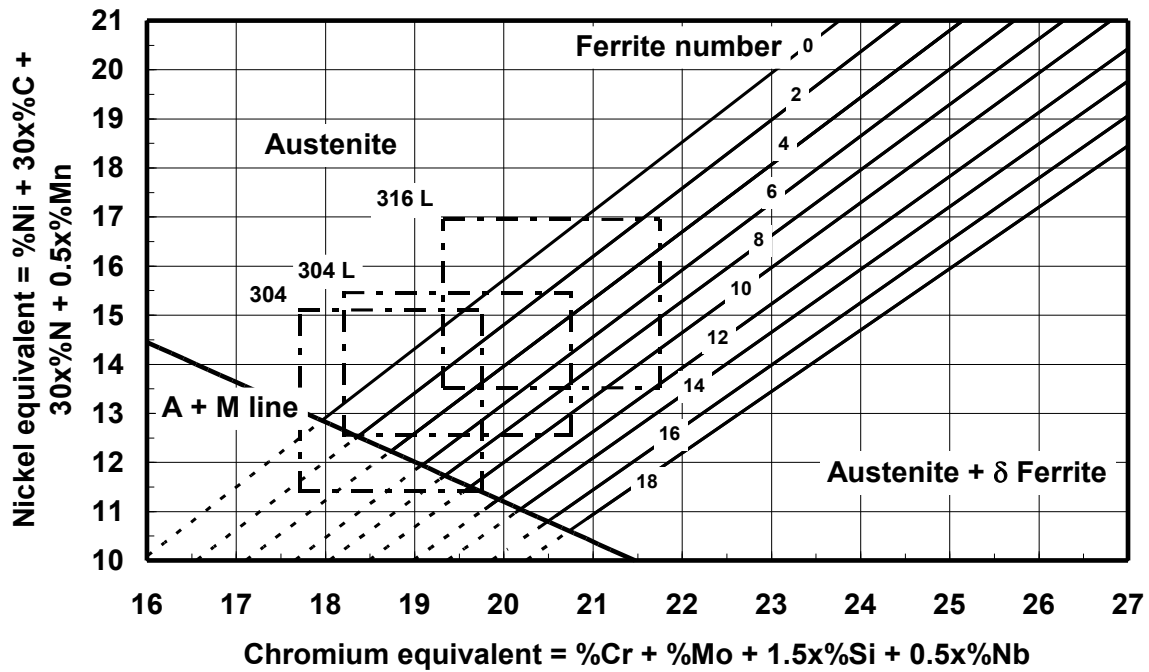

**Fig. 11:** Diagram for the determination of the ferrite content in austenitic stainless steel weld metal according to DeLong [50,51]. Diagrams more adapted to high N stainless steels are the ones of Hull [52] and Espy [53].

*3.5.1.2  Minimizing the risk of microfissuring in fully austenitic stainless steel welds*

Figure 12 shows a so-called 'Suutala diagram' allowing the risk of hot cracking to be predicted, as a function of the solidification mode (primary ferrite or austenite) and the impurity content of the steel (P, S). The primary mode of solidification, be it austenite or ferrite, is important for predicting the integrity of the weld. The ratio of $Cr_{eq}$ to $Ni_{eq}$ identifies the two modes in the diagram, primary ferrite for $Cr_{eq}/Ni_e > 1.5$ approx., primary austenite for $Cr_{eq}/Ni_e < 1.5$. In the domain of solidification to primary austenite, only the grades, fillers or combination of them corresponding to a very limited total residual element content can be considered safe in terms of hot cracking. Particular care should be taken when mixing base materials of very different origin and quality. Austenitic stainless steels such as 316LN properly specified [27−29] generally solidify in the fully austenitic range and show limited impurity content. If mixed with a general purpose stainless steel such as 304L or 316L, or so called 'free machining' grades with added S to improve machining rates, a risk of cracking cannot be excluded depending on the dilution. Austenitic stainless steels with high impurity contents are generally designed to solidify in the primary δ-ferrite range, in order to avoid risks of hot cracking. Their dilution with a fully austenitic grade might locally result in an austenitic weld with equivalent unacceptable level of impurities.

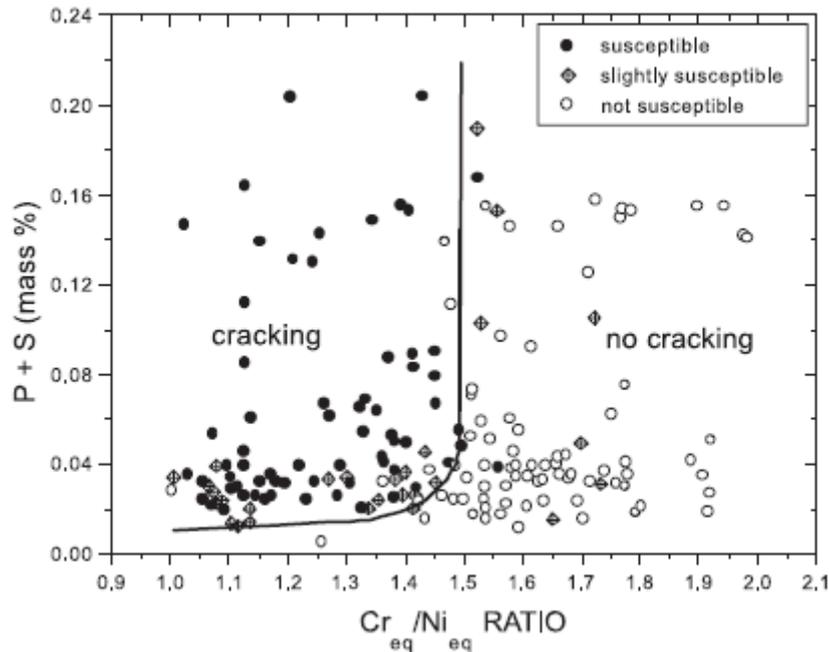

**Fig. 12:** Cracking susceptibility during arc welding of austenitic stainless steels ('Suutala' type diagram [54−57]). Equivalents based on Schaeffler equivalent formulae for $Cr_{eq}$ and $Ni_{eq}$, $Cr_{eq} = Cr + 1.5Si + 1.37Mo$, $Ni_{eq} = Ni + 0.31Mn + 22C + 14.2N$. Variants of the diagram exist to take also into account the effect of B.

Due to this risk, the use of free-machining grades, where addition of 150 ppm to 300 ppm S is allowed [30], should be strictly avoided. Components to be welded together in the vacuum system of an accelerator can be of very different origin and supplies. Their impurity content and solidification mode should be carefully checked to avoid leaks in the welds due to microfissuring.

### 3.5.2　*Solution annealing, outgassing treatments (vacuum firing), stress relieving*

Austenitic stainless steels are to be supplied and preferentially used in their solution annealed condition. Solution annealing consists of a heat treatment in a suitable temperature and time range adapted to the specific grade and size of the product, followed by quenching in water or rapid cooling by other means. Solution annealing allows optimum ductility and formability, toughness, and corrosion resistance to be achieved. Consistently, standards in force generally require the supply of products in the solution annealed condition, except for specific applications. Solution annealing is performed according to annealing requirements imposed by standards, at temperatures generally above 1040 °C to avoid sensitisation and precipitation of secondary phases. Solution annealing also allows ductility of cold finished products such as cold rolled sheets or drawn tubes to be recovered and hardness to be reduced. The maximum allowed hardness of products is limited by corrosion standards for use in severe environments [58]. CERN specifications also limit maximum allowed hardness.

Post-weld heat treatments are generally not required for austenitic stainless steels. On the other hand, for vacuum applications, a so-called 'vacuum firing' of components and subassemblies might be needed to outgas the material, in order to "remove the dissolved gas load in cleaned and degreased parts" [59]. 316LN products purchased to CERN specifications are compatible with applications requiring vacuum firing at 950 °C. Even if 950 °C is substantially below the solution annealing temperatures for austenitic stainless steels, a short treatment for a few hours at this temperature has no measurable detrimental metallurgical effects, owing to the presence of N in the grade that delays sensitisation.

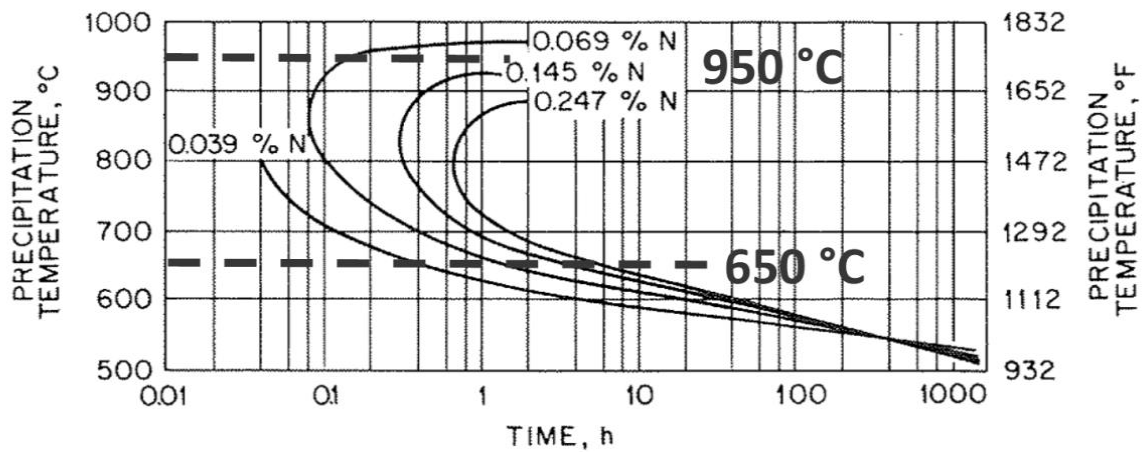

**Fig. 13:** Effect of N on precipitation of carbides ($M_{23}C_6$) in a 317 type stainless steel (0.05%C, 17%Cr, 13%Ni, 5%Mo) whose composition is close to 316LN. The N content range specified by CERN for 316LN is 0.14 % to 0.20 %. For such content, a treatment at 950 °C in the few hours range is expected to have no metallurgical effects, while treating at 650 °C for e.g., 24 h might provoke intergranular precipitation of carbides and sensitisation (from Refs. [17] and [60]).

On the other hand, products in 316L (a grade nominally not destined to be vacuum fired) might also be submitted to vacuum firing for specific applications. Moreover, for 316LN products destined to be coated, vacuum firing at 950 °C might be detrimental in function of the B content of the grade, and alternative treatments at 650 °C in the tens of hour range are applied[3]. These treatments might result in metallurgical effects. Figure 13 shows the effect of holding time and temperature on a N bearing stainless steel (composition close to EN 1.4439 or 317LN, a grade similar to 316LN), as a function of the N content. For a standard vacuum firing at 950 °C, no intergranular precipitation of carbides ($M_{23}C_6$) is expected for a few hour holding. This conclusion takes into account the specified N ranges of CERN (0.14% to 0.20% [28]) or standard (0.12% to 0.22% [30]) 316LN products and the low C content of 316LN (max. 0.030%). On the other hand, a treatment at 650 °C for prolonged times at temperature induces metallurgical effects that should be carefully assessed.

Figure 14 shows the detrimental effect of a heat treatment at 650 °C and 200 h (to mimic the reaction heat treatment of $Nb_3Sn$) on the ductility at cryogenic temperature of a jacket specimen of the ITER Toroidal Field (TF) conductor in modified 316LN IG (ITER grade). When tensile tested at 7 K following a light cold work (compaction and 2.5% stretching to simulate bending and straightening as undergone by the jacket during coil winding) and the above heat treatment, the specimen featured a ductility of only 11%, in spite of a severe specification limiting the C content to 0.02% max. (with a target < 0.015% against sensitisation). A tight control of chemical composition and manufacturing processes eventually allowed an elongation at breakdown of at least 20% to be maintained [62]. For comparison, standard 316LN grade in the solution annealed condition can feature ductility at 4.2 K above 40%.

---

[3] For a B content in solid solution above 8 ppm, vacuum firing of 316LN carried out at 950°C induces formation of BN at the surface [61], which is detrimental for the adherence of coated thin films and might require an eventual electropolishing prior to coating.

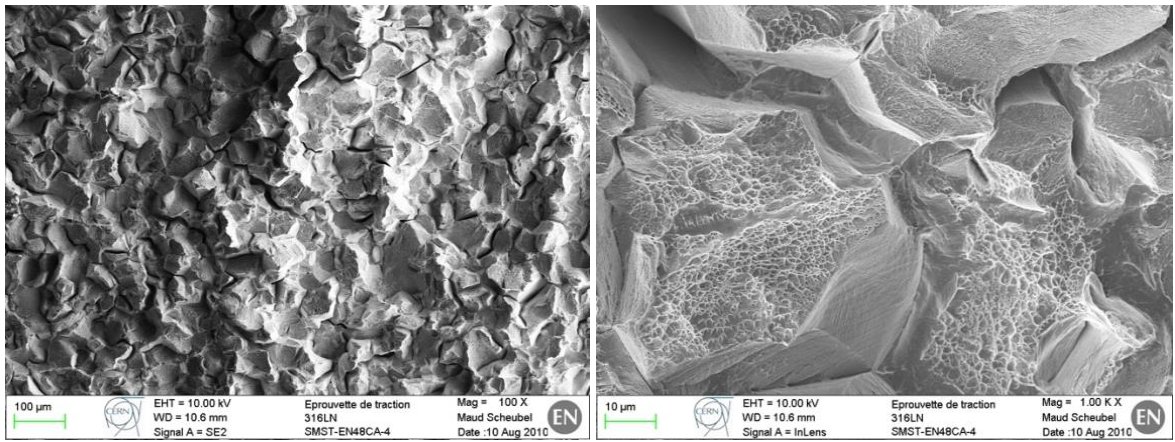

**Fig. 14:** Scanning Electron Microscope (SEM) fractographic analysis of a compacted and aged 316LN IG specimen of the jacket of the ITER toroidal field (TF) conductor tensile tested at 7 K, having been sensitised by a 650 °C – 200 h treatment following a light cold working. The observations confirmed quasi-cleavage with mainly intergranular fracture at the grain boundaries and limited dimpled areas within the single grains [63].

Stress relieving should also be preferentially operated outside the sensitising temperature range, either at low temperatures (i.e., under 500 °C for 316LN) or at 950 °C or above, where it can be made coincident with the 950 °C vacuum firing treatment. Additional restrictions might apply to welded structures, where σ-phase precipitation inducing detrimental effect on toughness can occur even above 950 °C. This is particularly relevant for Mo bearing grades and if the welds contain δ-ferrite, depending upon the ferrite content of the weld metal [51].

Sensitisation implies loss of corrosion resistance (Cr depletion at grain boundaries), loss of ductility (especially at cryogenic temperatures) with a ductile-to-brittle transition onset. The effect of a thermal treatment and the possible occurrence of sensitisation in austenitic stainless steels can be checked with the support of ASTM A262 standard, in particular through practice A and E [64].

### 3.6 Example of failures due to corrosion of stainless steels components [65]

Corrosion phenomena do not only occur during the operation or maintenance of stainless steel components; they often occur at the fabrication, testing or storage stages of parts or assemblies, too. In recent years, major corrosion failures have occurred in both accelerator and fusion magnet systems. One example is the corrosion failure of the ITER thermal shields (TS), that are the main barrier to minimise the transfer to the heat load to magnets operating at 4.5 K. This transfer is due to the radiation of warm components, in particular the cryostat and the Vacuum Vessel (VV) [66]. The TS consists mainly of silver-coated AISI 304LN stainless steel panels, which are cooled by the circulation of helium gas at 80 K, pressurised to 1.8 MPa in AISI 304L DN8 tubes stitch welded to the panels. The piping system totals 23 kilometres. SFA 5.9 ER308L was used as the filler material for the Tungsten Inert Gas (TIG) fillet stitch welding. Silver plating, a highly acidic process involving HCl-based surface activation, was applied to the welded panels. After the panels were delivered to IO and assembly began, three delayed leaks were discovered in 2021. These were associated with corrosion spots and residues, and were all located at the start of the stitch weld between the panel and the pipe. This is an area featuring high tensile residual stress and secluded volumes where Cl residues were retained. These residues, which arose from the early stages of the coating process, were not efficiently removed by subsequent water rinsing. The failure analysis [66, 67] concluded that Stress Corrosion Cracking (SCC) events had occurred, which had developed progressively through the thickness of the welds, despite them being qualified to stringent level B of ISO 5817 (Fig. 15).

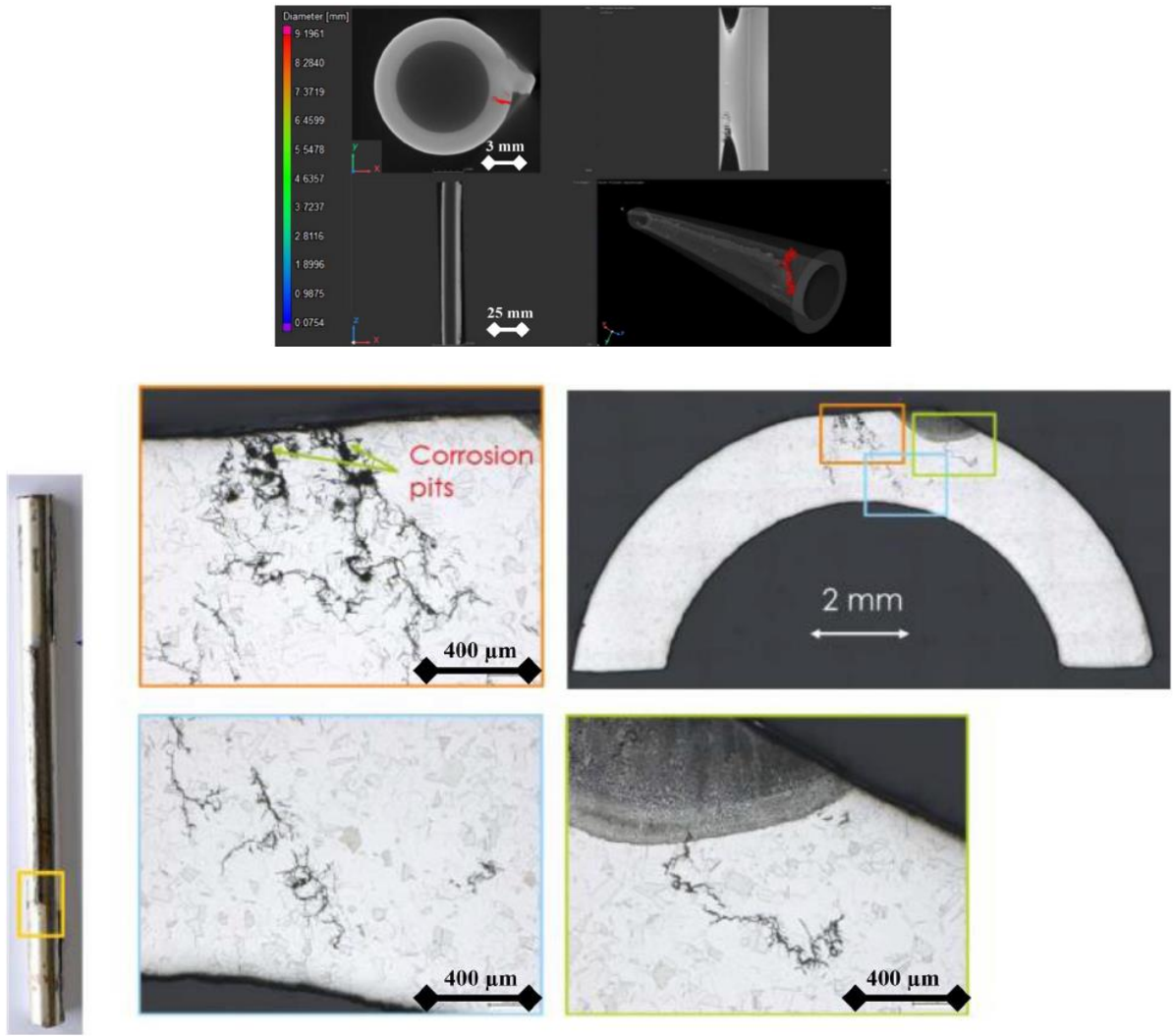

**Fig. 15**: Top – Computed Tomography (CT) scan of a shroud showing a branched crack crossing the entire cross-section of the tube. Bottom – Metallographic inspection of cross sections of a thermal shield. Transgranular, branched cracks merging from the outer wall and progressing towards the inner wall are observed. Pits are observed initiating at a distance of about 1 mm from the weld seam. Etching: 10% vol oxalic acid.

In order to prevent additional and eventual occurrence of SCC, the TS have been systematically repaired, including the ones of the already installed TF module. More specifically for both the Vacuum Vessel Thermal Shields (VVTS) and the Support Thermal Shields (STS), 100% of the cooling pipes have been removed, as well as a 2 mm thick layer under the welds to remove the Ag coating and a volume underneath possibly affected by SCC. Finally, NDT has been applied to exclude any corrosion events, including cracking in the panel. The installation of the new pipes (in the higher alloy grade 316L) has not been followed by Ag coating. Filler metal has been changed to ER 317L Mn modified, featuring a higher Pitting Resistance Equivalent Number (PREN). This has allowed to suppress the existing SCC events and remove the root cause of the observed leaks, as the source of chlorides contamination is eliminated where the new pipes have been welded (Fig. 16).

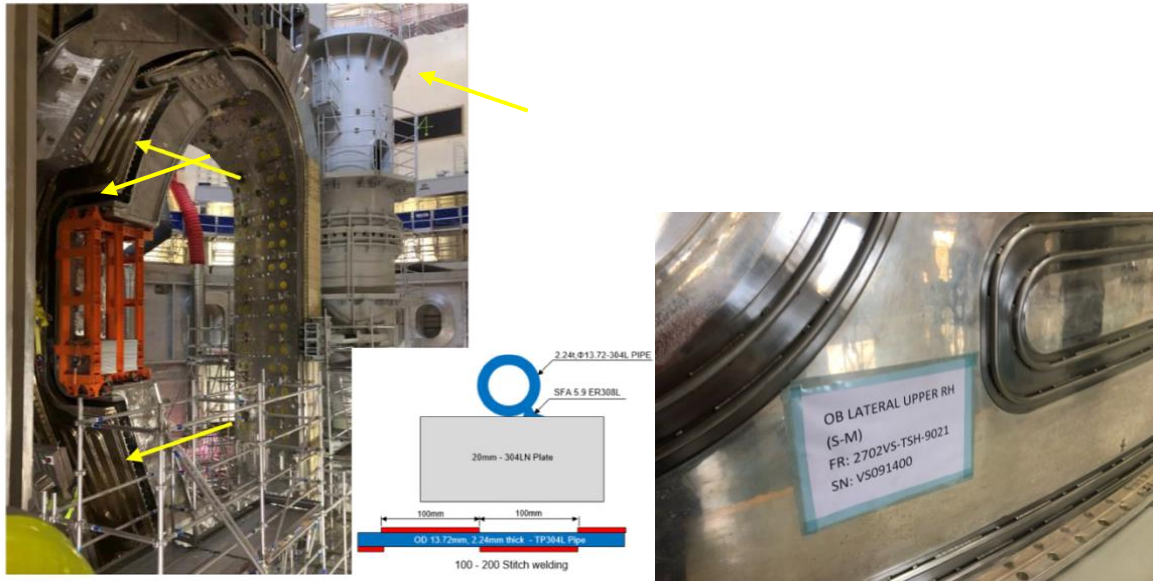

**Fig. 16**: Left - The first subassembly of the ITER VV in place within the tokamak pit. Pipes welded onto the surface of the TS panels are visible in this picture (identified by arrows). Middle - Pipe stitch welding configuration. The old 304L pipes have eventually been replaced by 316L ones and filler metal has been changed to ER 317L Mn modified for increased corrosion resistance. Right - Repaired VVTS panel with new pipes installed through stitch welding, after complete removal of the old cooling pipes from the panels, machining of the pipe path to suppress adjacent volumes possibly affected by onset of SCC, and stitch welding of new piping to the surface (yellow arrows). Almost eight months were needed for repairing one set of VVTS.

## 4 Conclusions

The description of several applications of steel and stainless steels for the construction of large particle accelerators and high-energy physics experiments, of fusion reactors and their superconducting magnet structures shows that a material not only consists of a 'chemical composition' or a designation, but is also the result of a complete metallurgy and metalworking process including possible refinement and remelting steps. A material for a demanding application is also defined through specified non-destructive and destructive tests, and should be ordered and delivered with a complete certification. In some cases, a quality control plan should be defined between the producer and the customer and should be followed during the production. The expected cost for a material adapted to an envisaged application will depend on an adequate specification defined before the order. Taking as an example austenitic stainless steels, a general purpose 304L, issued from a simple electric furnace primary melting, can have a cost that is 30 times lower than an ESR product multidirectionally and redundantly forged to a particular final size and properly specified for the final application.

The severity of the specification will depend on the final application. Whenever leak tightness has to be guaranteed across a thin wall, an adequate microstructure in terms of grain size and inclusion content has to be specified. A fabrication resulting in a proper texture (orientation of the fibres parallel to the walls, or if this is not achievable, absence of orientation through multidirectional forging) should be preferred.

In most constructions, several components have to be integrated through delicate welding operations, introducing internal stresses that can bring to leak a previously tested leak tight component. Advanced inspection techniques such as Phased Array Ultrasonic Testing (PAUT) or CT are available, extending the possibilities of conventional NDT. In particular, CT might allow for a 100% volumetric inspection of welds otherwise virtually non-inspectable by conventional NDT.

Applications at cryogenic temperatures, possibly associated with a requirement of non-magnetism, demand special care. A stainless steel, such as 316LN, specified as fully austenitic in order to be non-magnetic at RT will show, at cryogenic temperature, a higher magnetic susceptibility that might be unacceptable for some components. For the beam screen of LHC, a special austenitic stainless steel had to be developed to maintain low magnetic susceptibility in the base metal and the welds, in the temperature range between 10 K and 20 K. Strength and ductility requirements should not only be considered for the working temperature of a component, but defined as a function of all of the temperature cycles undergone by the component (vacuum firing, $Nb_3Sn$ reaction heat treatment, stress relieving, baking, etc. as applicable).

Proper selection of materials should take into account their availability (special grades of stainless steels might require more than one year of lead time for delivery).

Corrosion issues should not be neglected, especially for thin walled components. The use of halogen-activated fluxes should be strictly avoided in a stainless steel environment [68].

Material aspects should always be considered at the beginning of a project: issues of availability, interest in considering in a timely manner alternative techniques, including near-net shaping, as often late selections make the project more costly or incompatible with the tight schedule of an accelerator project. New projects will eventually result in less conservative solutions (bimetals by HIP assisted diffusion bonding or explosion bonding, additive manufacturing).